\documentclass[%
reprint,
superscriptaddress,
 amsmath,amssymb,
prb,
]{revtex4-2}

\usepackage{graphicx}
\usepackage{dcolumn}
\usepackage{bm}
\usepackage{hyperref}

\begin{document}

\preprint{APS/123-QED}

\title{Spin polarisation and spin dependent scattering of holes in transverse magnetic focussing}
\author{M. J. Rendell}
\thanks{M. J. Rendell and S. D. Liles contributed equally to this work}
\affiliation{
School of Physics and Australian Research Council Centre of Excellence in Future Low-Energy Electronics Technologies, University of New South Wales, Sydney, NSW 2052, Australia
}
\author{S. D. Liles}
\author{A. Srinivasan}
\affiliation{
School of Physics, University of New South Wales, Sydney, NSW 2052, Australia
}
\author{O. Klochan}
\affiliation{
University of New South Wales Canberra, Canberra, ACT 2600, Australia
}
\affiliation{
School of Physics and Australian Research Council Centre of Excellence in Future Low-Energy Electronics Technologies, University of New South Wales, Sydney, NSW 2052, Australia
}
\author{I. Farrer}
\affiliation{
Department of Electronic and Electrical Engineering, University of Sheffield, Sheffield, S1 3JD, UK
}
\affiliation{
Cavendish Laboratory, University of Cambridge, Cambridge, CB3 0HE, UK
}
\author{D. A. Ritchie}
\affiliation{
Cavendish Laboratory, University of Cambridge, Cambridge, CB3 0HE, UK
}
\author{A. R. Hamilton}
\email{alex.hamilton@unsw.edu.au}
\affiliation{
School of Physics and Australian Research Council Centre of Excellence in Future Low-Energy Electronics Technologies, University of New South Wales, Sydney, NSW 2052, Australia
}
\date{\today}

\begin{abstract}
In 2D systems with a spin-orbit interaction, magnetic focussing can be used to create a spatial separation of particles with different spin. Here we measure hole magnetic focussing for two different magnitudes of the Rashba spin-orbit interaction. We find that when the Rashba spin-orbit magnitude is large there is significant attenuation of one of the focussing peaks, which is conventionally associated with a change in the spin polarisation. We instead show that in hole systems with a $k^3$ spin-orbit interaction, this peak suppression is due to a change in the scattering of one spin state, not a change in spin polarisation. We also show that the change in scattering length extracted from magnetic focussing is consistent with results obtained from measurements of Shubnikov-de Haas oscillations. This result suggests that scattering must be considered when relating focussing peak amplitude to spin polarisation in hole systems.
\end{abstract}

\maketitle


\section{\label{sec:Intro}Introduction}

In a magnetic focussing experiment, a collimated beam of charge is focussed into a circular orbit by a transverse magnetic field, analogous to a mass spectrometer. Magnetic focussing was originally proposed as a method of studying the Fermi surface of metals \cite{sharvin_possible_1965, tsoi_focusing_1974}, and has also been used to measure band structures in graphene \cite{taychatanapat_electrically_2013}, and electron-electron scattering lengths in GaAs/AlGaAs \cite{gupta_precision_2021}. 

In systems with a spin-orbit interaction (SOI), the magnetic focussing trajectories become spin dependent as the spin states are now coupled to momentum. If the SOI is sufficiently large, magnetic focussing can spatially separate the spin states and create a spin-dependent mass spectrometer \cite{rokhinson_spin_2004, dedigama_current_2006, rokhinson_spontaneous_2006, heremans_spin-dependent_2007, zulicke_magnetic_2007, bladwell_magnetic_2015, lo_controlled_2017, lee_influence_2022}. The high mobility and large SOI of 2D hole systems in GaAs has made them an ideal candidate for spin-dependent magnetic focussing experiments. Experimental work has used magnetic focussing to measure spatial separation of spin \cite{rokhinson_spin_2004}, spin filtering by quantum point contacts (QPCs) \cite{rokhinson_spontaneous_2006} and interactions between 1D subbands in a QPC \cite{chesi_anomalous_2011}. Magnetic focussing of holes has also been proposed as a way to measure g-factor anisotropies \cite{bladwell_measuring_2019}, and complex spin dynamics \cite{zulicke_magnetic_2007, bladwell_interference_2018}, which are not visible in other measurements of 2D systems such as Shubnikov-de Haas oscillations.

Here, we concentrate on the use of magnetic focussing peak amplitude as a measure of the spin polarisation \cite{potok_detecting_2002,folk_gate-controlled_2003,rokhinson_spin_2004,rokhinson_spontaneous_2006,chen_all-electrical_2012}. It has been proposed that the relative amplitudes of the spin-split magnetic focussing peaks is determined by the spin polarisation of the injected charge. This technique has been used in hole systems to observe spontaneous polarisation in QPC transmission \cite{rokhinson_spontaneous_2006} and spin-dependent transmission of QPCs \cite{rokhinson_spin_2004, chesi_anomalous_2011}. Despite magnetic focussing being used for these techniques, there has been limited study of the effect of changing the magnitude of the Rashba SOI on hole magnetic focussing.

A recent study investigated magnetic focussing using a device where the Rashba SOI magnitude could be tuned in situ using a top gate voltage (V$_{\text{TG}}$) \cite{rendell_gate_2022}. This technique revealed an increase in the spatial separation of the spin-split focussing trajectories as the Rashba SOI was increased. However, there is a limit to the amount the Rashba SOI can be changed using this method. In addition, any change to $V_{TG}$ will also change the 2D hole density and confining potential in addition to the Rashba SOI magnitude. As such, further study requires a different method of changing the Rashba SOI.

In this work we study magnetic focussing in two lithographically identical samples which differ only in the magnitude of the Rashba SOI. We change the Rashba SOI by changing the heterostructure used to confine the 2D system, allowing us to create a large change in the magnitude of the Rashba SOI for a similar V$_{\text{TG}}$ and 2D density. By comparing the two samples, we observe a change in the amplitude of the magnetic focussing peaks, which is typically associated with a change in the spin polarisation. However, we instead find that the change in peak amplitude is consistent with an increase in scattering of one spin state rather than a change in spin polarisation. We measure the scattering length of each spin state from the focussing peak amplitude, and find good agreement with scattering lenghts found from Shubnikov-de Haas measurements. We conclude that the change in focussing peak amplitude is due to the $k^3$ Rashba term causing a different effective mass and hence scattering length of each spin state, rather than a change in spin polarisation. This result suggests that care must be taken when relating the amplitude of spin-split focussing peaks to the spin polarisation in 2D hole systems.

\section{\label{sec:HolesVsElectrons}Magnetic focussing with a cubic Rashba spin-orbit interaction}

\begin{figure}
    \centering
    \includegraphics[width=0.45\textwidth]{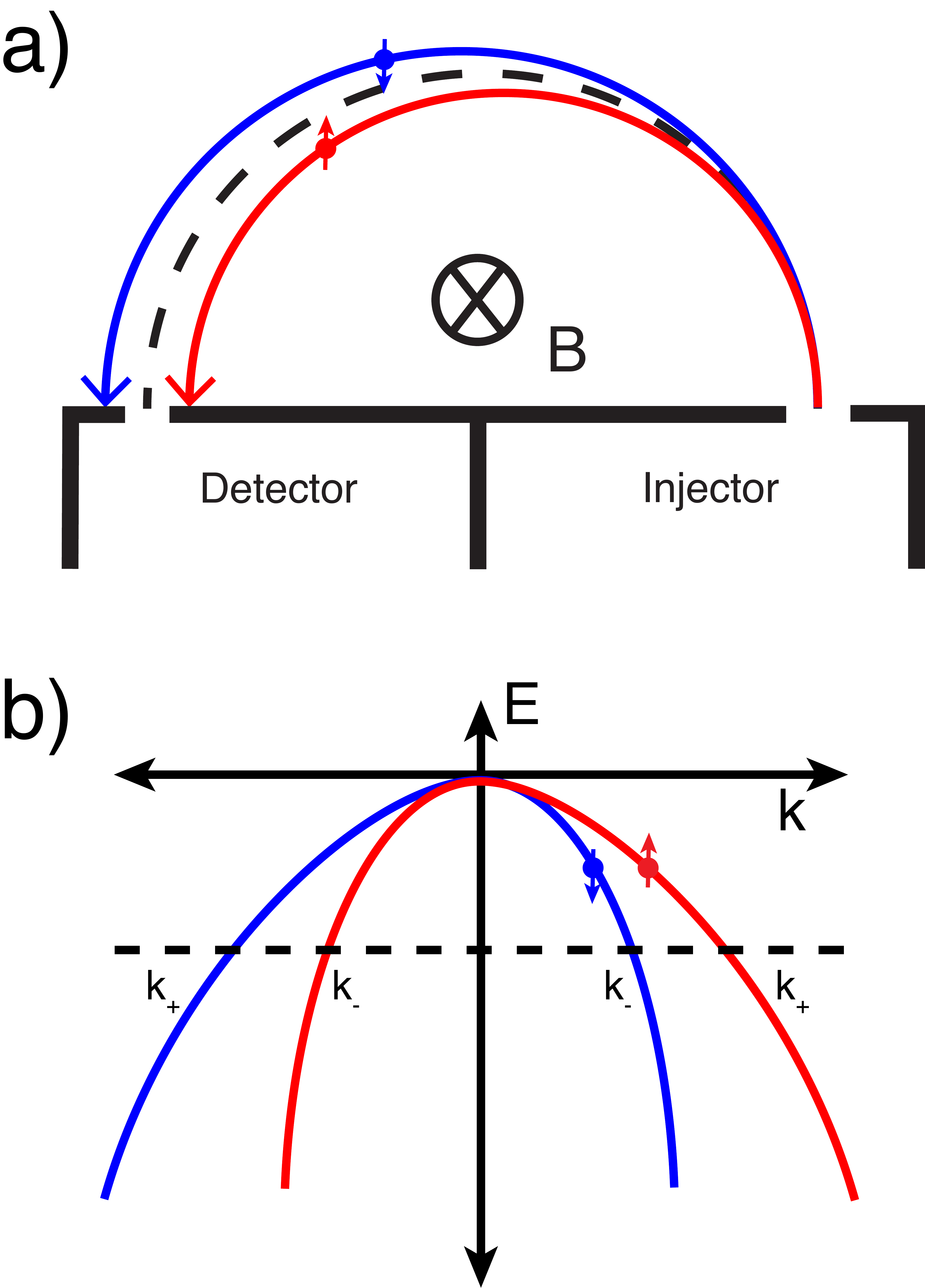}
    \caption{\textbf{a)} Magnetic focussing in the presence of a spin-orbit interaction. The red and blue lines correspond to the spin split focussing trajectories, which result in a splitting of the first focussing peak. The dashed line corresponds to the classical focussing trajectory. \textbf{c)} The first 2D subband for a hole system. Here the Rashba SOI term depends on $k^3$ which causes a change in the slope (and hence $m^*$) of the spin resolved subbands.}
    \label{fig:FocussingSchematic}
\end{figure}

Fig \ref{fig:FocussingSchematic}a) shows a schematic of a hole magnetic focussing device. A constant current is applied through an injector, where an out-of-plane perpendicular magnetic field causes the holes to follow cyclotron orbits. The voltage across the detector is measured, and a peak is observed when the focussing diameter is equal to the spacing between injector and detector (black dashed line in Fig \ref{fig:FocussingSchematic}a). Peaks in the focussing signal occur when the magnetic field is an integer multiple of \cite{van_houten_coherent_1989}
\begin{equation*}
    B = \frac{2\hbar k_F}{e d}
\end{equation*}
Where $k_\text{F}$ is the Fermi momentum and $d$ is the distance between injector and collector QPC (focussing diameter). In the presence of a spin-orbit interaction (SOI) the hole trajectories become spin dependent, resulting in a spatial separation of spin (blue and red lines in Fig \ref{fig:FocussingSchematic}a). The spatial spin separation causes the first magnetic focussing peak to split into two, with each peak corresponding to a different spin chirality. The relative amplitude of these spin peaks has been used as a measure of the spin polarisation in 2D hole systems.\cite{rokhinson_spin_2004}

The form of the Rashba spin-orbit term for 2D hole systems is fundamentally different to equivalent electron systems. This difference can have a dramatic impact on spin resolved focussing peaks. In GaAs, the subband dispersion for 2D holes with a Rashba SOI is given by \cite{winkler_spin-orbit_2003}

\begin{equation}\label{eq:HoleRashba}
    \mathcal{E}_{\text{h}} = \frac{\hbar^2 k^2}{2m^*} \pm \frac{\beta E_\text{z}}{\Delta_{\text{HH-LH}}} k^3
\end{equation}

where $E_\text{z}$ is the electric field in the out-of-plane direction and $\Delta_{\text{HH-LH}}$ is the splitting between the heavy hole (HH) and light hole (LH) subbands. Fig \ref{fig:FocussingSchematic}b) shows the resulting HH subband dispersion for a 2D hole system with Rashba SOI. The SOI causes the momentum of the holes to become spin dependent, with two values of $k$ ($k_+$ and $k_-$) at the Fermi energy (horizontal dashed line). In a magnetic focussing measurement, this results in separate cyclotron orbits for each spin and creates a spatial spin separation, splitting the first focussing peak. Previous work has demonstrated the ability to detect a change in peak splitting as the magnitude of the Rashba SOI is changed \cite{rendell_gate_2022}.

The $k^3$ structure of the Rashba SOI term for holes also causes the curvature of the 2D subbands to become spin dependent. This results in a difference in effective mass for each spin chirality in addition to the difference in $k$ \cite{stormer_energy_1983}. The spin dependent effective mass has been used to demonstrate electrical control of the Zeeman splitting \cite{marcellina_electrical_2018}, and proposed as a way to detect and generate topological properties in a 2D hole system \cite{marcellina_signatures_2020, cullen_generating_2021}. The change in effective mass is also possible to detect via focussing peaks. If the Rashba SOI term is sufficiently large, the difference in effective mass can be observed as a difference in scattering. Since focussing peak amplitude is exponentially sensitive to scattering \cite{heremans_observation_1992, rendell_transverse_2015}, the change in effective mass will therefore impact the focussing peak amplitude. This analysis does not include contributions from k-linear Rashba SOI terms for 2D holes \cite{rashba_spin-orbital_1988, luo_discovery_2010, durnev_spin-orbit_2014}. These terms do not cause a spin-dependent change in the curvature of the 2D subbands and should not affect the difference in effective mass between the spin chiralities.

\section{\label{sec:Interface}Interface dependence}

To create a large change in the Rashba SOI without changing the carrier density, we prepare two samples with different interface symmetry. One sample uses a 15nm (100) GaAs/Al$_{0.33}$GaAs$_{0.67}$ quantum well (QW) heterostructure with a square well like 2D confining potential (wafer W713). The other is a (100) GaAs/Al$_{0.33}$GaAs$_{0.67}$ single heterojunction (SHJ) which creates a triangular confining potential (wafer W640). Changing from a QW to a SHJ reduces the 2D confinement, decreasing $\Delta_{\text{HH-LH}}$ without causing a large change in $k$ and $E_\text{z}$. From Eq \ref{eq:HoleRashba}, the SHJ device (smaller $\Delta_{\text{HH-LH}}$) will have a larger Rashba SOI term and therefore larger focussing peak splitting. The SHJ device will also have a larger difference in effective mass between the spin subbands.

\begin{figure*}
    \centering
    \includegraphics[width=\textwidth]{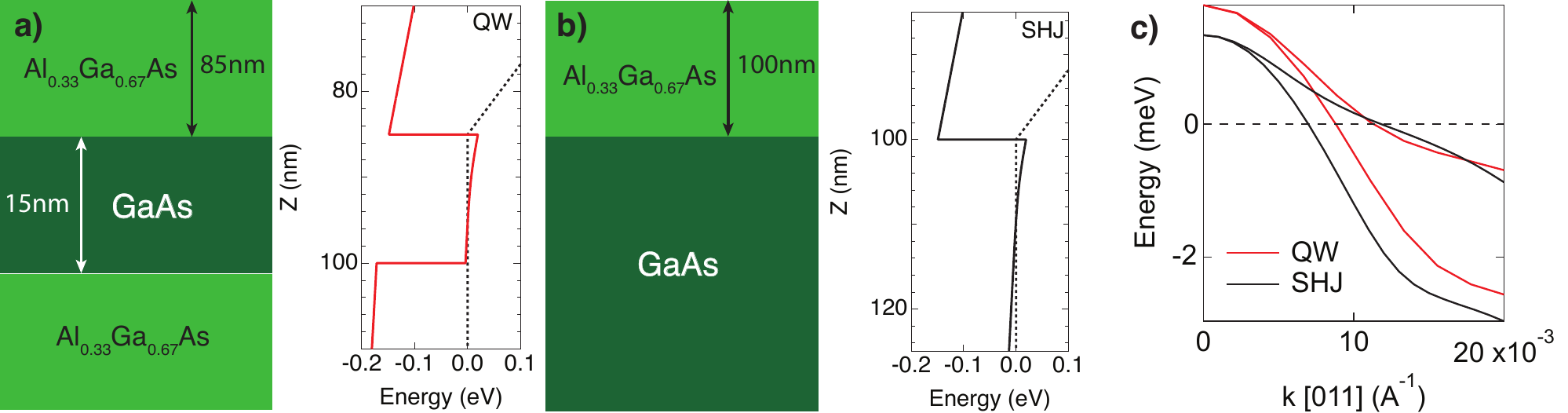}
    \caption{Comparison of sample heterostructures. \textbf{a)} Heterostructure of the QW sample and resultant band edge. \textbf{b)} Heterostructure and band edge of the SHJ sample. \textbf{c)} A comparison of the HH1 subband dispersion for both samples. Calculated with a 6x6 k.p solver (Nextnano).}
    \label{fig:HeterostructureComparison}
\end{figure*}

Fig \ref{fig:HeterostructureComparison} compares the wafer structure and resultant confining potentials for the QW (a) and SHJ (b) samples. The left side of each panel is the wafer structure around the 2D interface, while the right side shows the resulting band edge found using a Schrodinger-Poisson solver (Nextnano \footnote{\url{http://www.nextnano.de/}}). The $E(k)$ dispersion relations of both samples are also calculated using Nextnano. This calculation uses a $6\times 6$ $k\cdot p$ solver and includes contributions from Rashba SOI terms but does not include Dresselhaus SOI terms. Fig \ref{fig:HeterostructureComparison} c) shows the spin split HH1 subbands for both samples, with a clear difference in $k$ between the HH+ and HH- subbands at $E = 0$ (horizontal dashed line). It is this difference in $k$ that results in a splitting of the first focussing peak in both samples.

There is a significantly larger splitting visible for the SHJ sample at $E_\text{F}$ (horizontal dashed line), which leads to an increase in the focussing peak spacing. There is also a large difference in the curvature of the HH+ and HH- subbands. The difference in curvature of the $E(k)$ dispersion results in a spin dependent effective mass, which can also be detected in a focussing measurement.

\begin{figure}
	\includegraphics[width=0.48\textwidth]{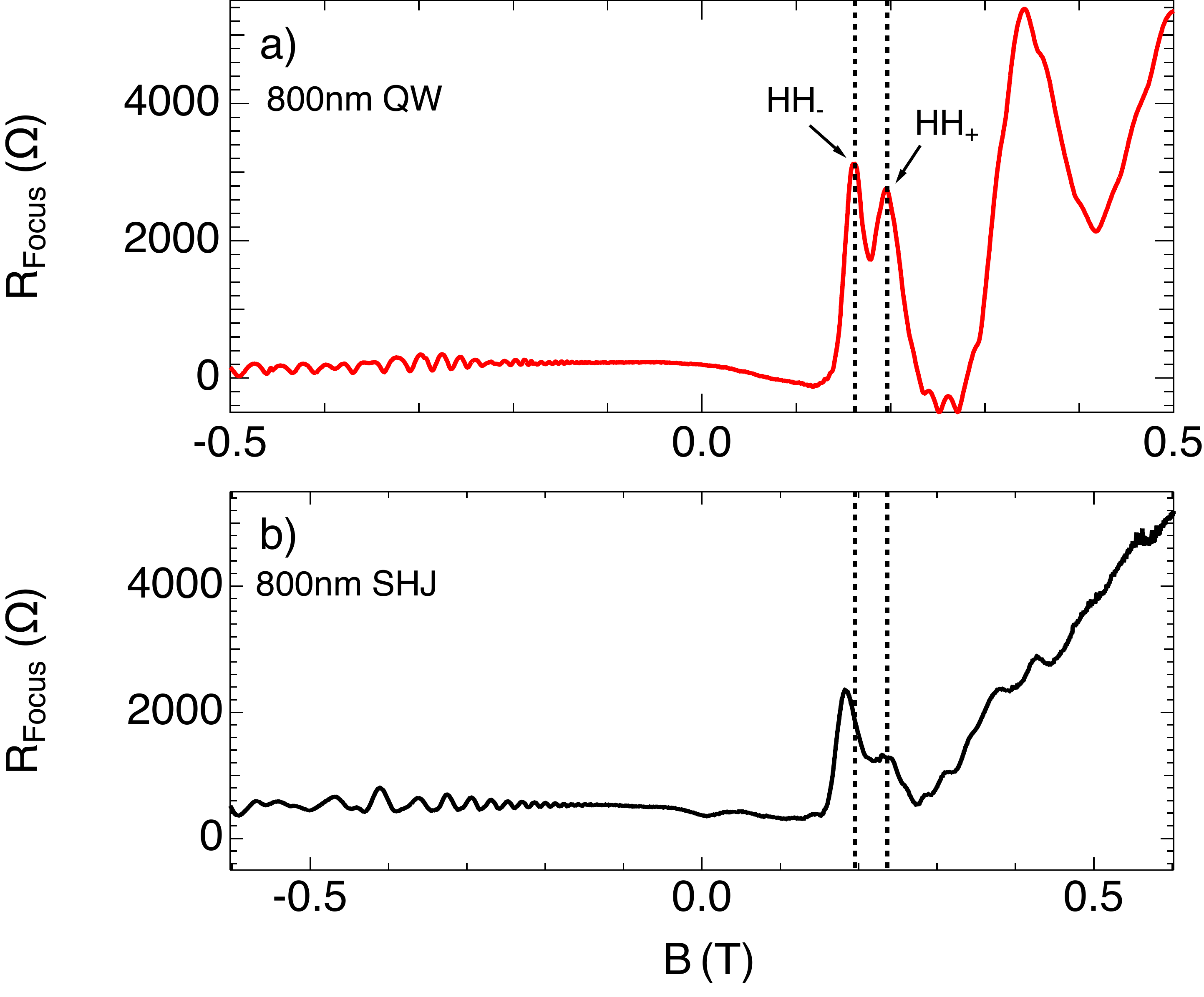}
    \caption{Spin-split magnetic focussing with different interface symmetry. Red is focussing in a 15nm quantum well with weaker Rashba spin-orbit ($n_{\text{2D}}$ = 1.57$\times 10^{11}$ cm$^{-2}$). Black is focussing in a single heterojunction with stronger Rashba spin orbit at the same $V_{\text{TG}}$ ($n_{\text{2D}} = 1.89\times10^{11} \text{cm}^{-2}$). Dashed vertical lines indicate the spacing of the 800nm QW focussing peaks. X-axis in the bottom panel has been scaled by the ratio of $n_{\text{2D}}$ to account for the difference in density.}
   \label{fig:RashbaDependence}
\end{figure}

Fig \ref{fig:RashbaDependence} compares focussing in the QW and SHJ samples over the same focussing diameter (800 nm). Starting with the QW sample (Fig \ref{fig:RashbaDependence}a), we observe a clear spin split focussing peak in positive $B$, with higher order peaks also observed. No splitting of higher order peaks is observed due to spin-flip reflections from the boundary \cite{reynoso_magnetic_2008, lee_influence_2022}. Vertical dashed lines indicate the position of the spin-split first focussing peaks. In the opposite $B$ polarity the holes are focussed away from the collector QPC and only Shubnikov-de Haas oscillations are visible.

The same measurement is repeated on a lithographically identical sample, fabricated on the SHJ wafer as shown in Fig \ref{fig:RashbaDependence}b. This measurement was performed at the same $V_{\text{TG}}$ as the QW sample, resulting in a slightly higher hole density ($n_{\text{2D}}$ = $1.89\times 10^{11} \text{cm}^{-2}$ vs $1.57\times 10^{11} \text{cm}^{-2}$). The increased Rashba SOI in the SHJ sample results in focussing peaks which are further apart than the QW sample. The focussing peaks in Fig \ref{fig:RashbaDependence}b) are also significantly smaller in amplitude than those in the QW sample, with the higher field peak attenuated and broader compared to the lower field peak. Typically such a difference in amplitude of spin-resolved focussing peaks is interpreted as a change in the spin polarisation \cite{rokhinson_spin_2004, rokhinson_spontaneous_2006}. However, here the spin polarisation should be approximately equal as both QPCs are biased to the G = 2e$^2$/h plateau to transmit both spin states.

To determine if the change in peak amplitude in the SHJ sample is instead related to an increase in scattering, the decay of focussing peak amplitude is measured over a range of focussing diameters. The device geometry of both focussing samples allows focussing to be measured for a range of focussing diameters (d = 800, 2300 and 3100nm). By measuring the change in peak amplitude as a function of focussing diameter, the scattering length of each of the spin peaks can be found \cite{heremans_observation_1992, rendell_transverse_2015}.

\begin{figure}
    \centering
    \includegraphics[width=0.48\textwidth]{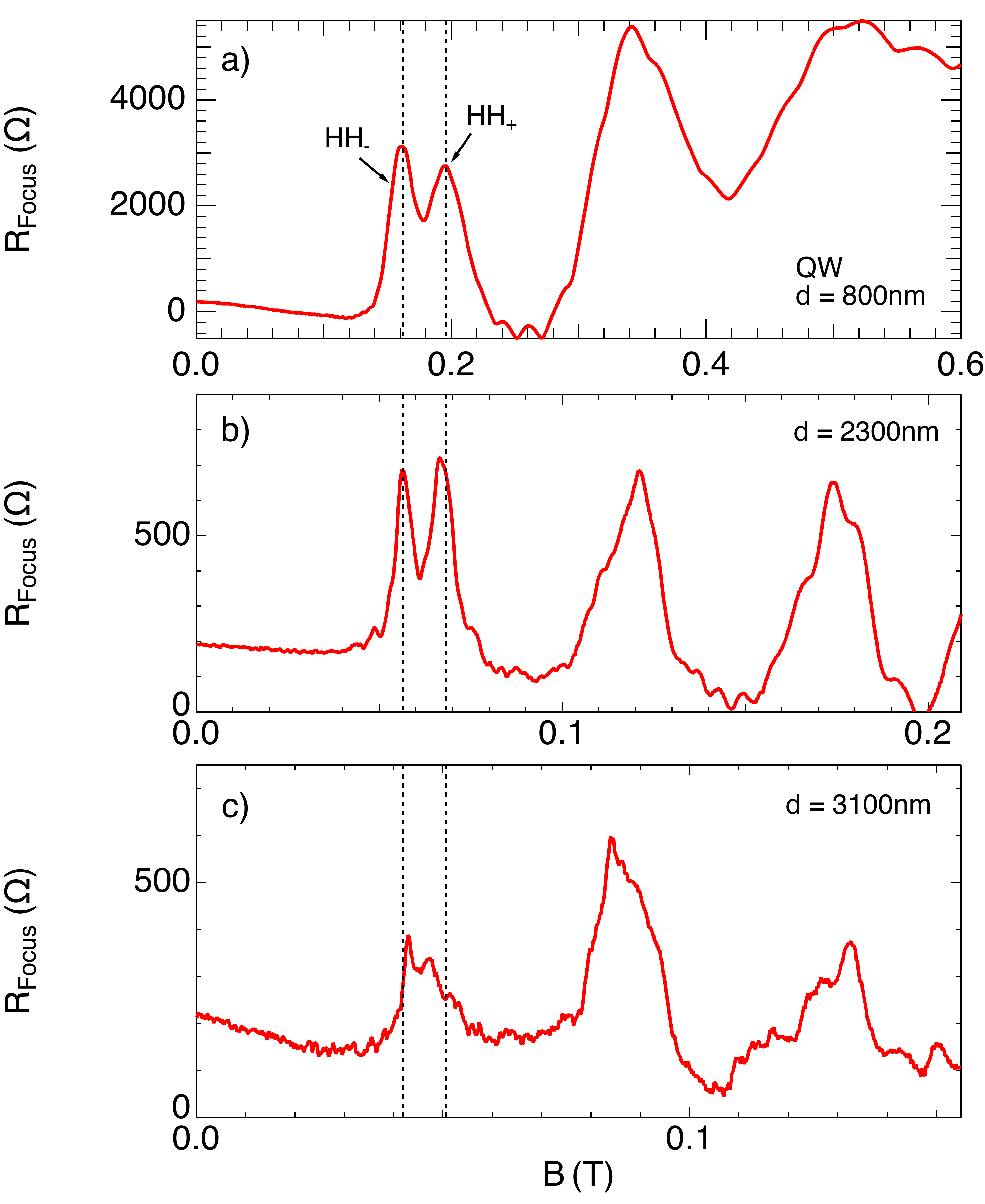}
    \caption{Focussing over different diameters in the QW sample. The focussing diameter is varied from 800nm (a) to 3100nm (c). The x-axis range of each panel has been chosen based on the focussing diameter. The vertical dashed lines indicate the position of the spin resolved focussing peaks in the 800nm trace.}
    \label{fig:QWlength}
\end{figure}

First, the focussing diameter dependence of the peak amplitude is measured on the QW sample. Fig \ref{fig:QWlength} shows focussing measured on the QW sample for all three focussing diameters. To allow for easy comparison between the focussing lengths, the $B$ axis range of each plot has been chosen based on the ratio of the focussing diameters. This should result in vertical alignment of the same focussing peaks across each diameter. Vertical dashed lines indicate the position of the spin split focussing peaks in Fig \ref{fig:QWlength}a) (focussing diameter = 800nm). As the focussing diameter is increased from 800nm (Fig \ref{fig:QWlength}a) to 2300nm (Fig \ref{fig:QWlength}b) there is good agreement in $B$ location between all of the peaks. Multiple higher order peaks can be observed, with the amplitude of the higher order peaks similar to the spin-split focussing peaks, indicating specular reflections from the boundary between the injector and detector QPCs. The spin split focussing peaks can also be clearly resolved, and align with the peaks in Fig \ref{fig:QWlength}a) as indicated by the vertical dashed lines. The spin split focussing peaks also have a similar amplitude, indicating an equal population of both spin states (i.e. no spin polarisation). As the focussing diameter is further increased to 3100nm, (Fig \ref{fig:QWlength}c) the effects of scattering begin to dominate the focussing signal. The amplitude of the first focussing peak is significantly reduced and both spin peaks can no longer clearly be resolved.

\begin{figure}
    \centering
    \includegraphics[width=0.48\textwidth]{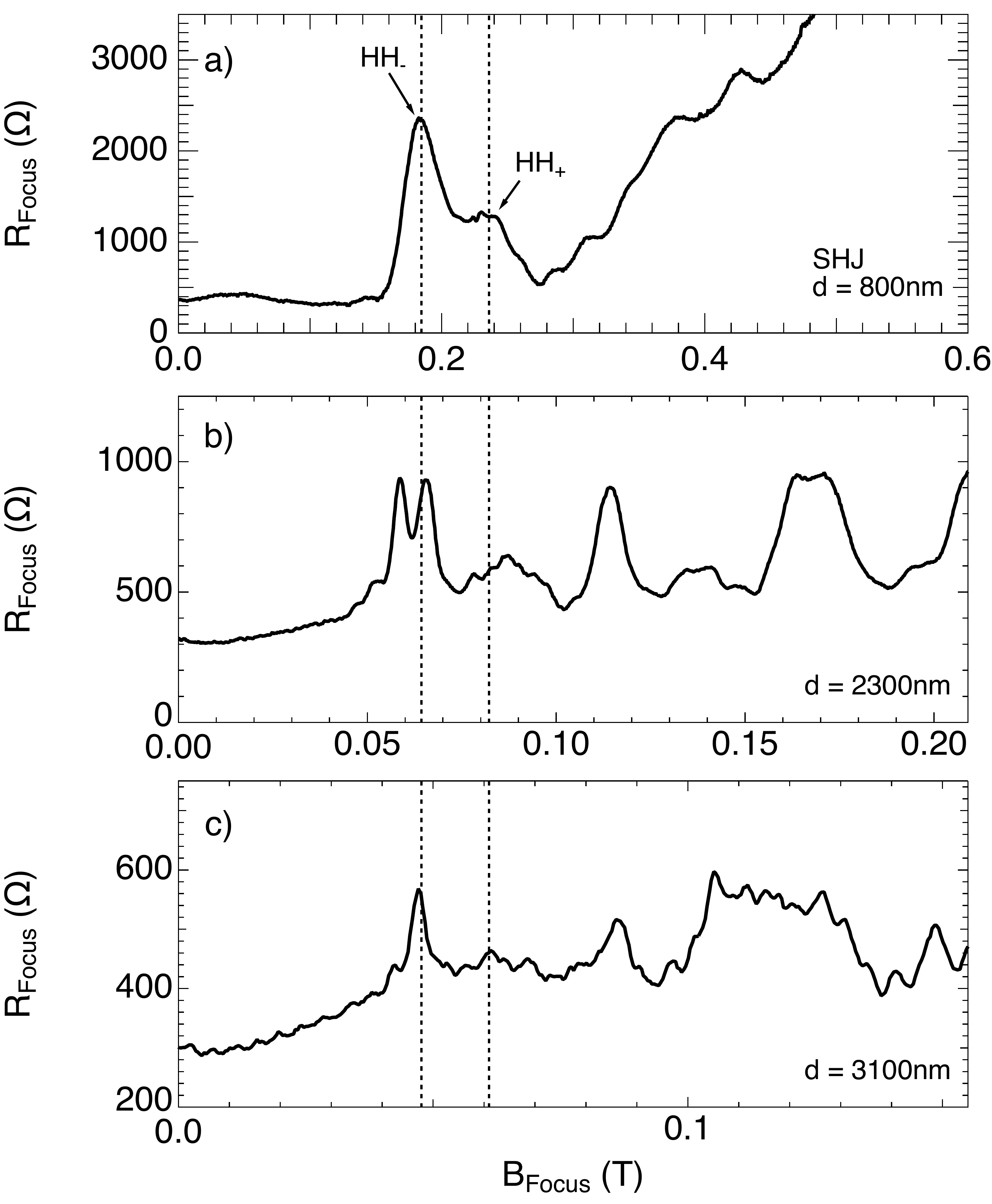}
    \caption{Focussing comparison for the SHJ sample. Focussing over a diameter of 800nm (a), 2300nm (b) and 3100nm (c). The x-axis range has been chosen to allow direct comparison of focussing peaks. Vertical dashed lines indicate the position of the spin resolved focussing peaks for the 800nm focussing diameter.}
    \label{fig:SHJlength}
\end{figure}

In Fig \ref{fig:SHJlength} focussing is measured in the SHJ sample for the same three focussing diameters (800, 2300 and 3100nm). Once again, the x-axis range of each panel has been chosen based on the focussing diameter, and the vertical dashed lines indicate the position of the spin-split focussing peaks in Fig \ref{fig:SHJlength}a). The HH- peak is narrow and large in amplitude, while the HH+ peak is significantly lower in amplitude and broader. The spacing of the peaks has also increased compared to the QW sample (Fig \ref{fig:QWlength}a), as expected for a larger Rashba SOI magnitude. As the focussing diameter is increased from 800nm (Fig \ref{fig:SHJlength}a) to 2300nm (Fig \ref{fig:SHJlength}b) multiple changes can be observed. First, the amplitude of all focussing peaks decreases with the increase in focussing path length. In particular, the HH+ spin peak becomes very broad and low in amplitude. In comparison, the HH- spin split peak is higher in amplitude and narrower, and both spin peaks have superimposed interference structure \cite{van_houten_coherent_1989, aidala_imaging_2007, bladwell_interference_2017, lee_influence_2022}. The HH- spin peak displays structure similar to a spin split focussing peak, however this splitting is too small to be caused by spin. The structure on the HH- peak is most likely due to an interference effect, as similar (but smaller amplitude) oscillations are visible on the low B side of the peak, a characteristic signature of interference due to diffraction in focussing \cite{bladwell_interference_2017, lee_influence_2022}. Finally, as the focussing diameter is increased to 3100nm (Fig \ref{fig:SHJlength}c) the amplitude of the HH+ and HH- peaks is further reduced, with the HH+ peak barely resolved due to scattering.

The significantly lower amplitude of the HH+ spin split peak in Fig \ref{fig:SHJlength} is consistent with the larger effective mass of this spin band (see Fig \ref{fig:HeterostructureComparison}c). Assuming the scattering time is the same, the larger effective mass should result in a shorter scattering length for the HH+ spin state. As the total path length travelled by both spin states is the same, being fixed by the focussing geometry, a shorter scattering length will result in more scattering for the HH+ state and hence a lower amplitude of the corresponding HH+ focussing peak. Since focussing measurements are exponentially sensitive to scattering effects \cite{spector_ballistic_1990,heremans_observation_1992,rendell_transverse_2015}, a change in scattering rate can be the dominant cause of the amplitude change, rather than a change in the spin polarisation.

\section{\label{sec:Scattering}Scattering length}

To understand the suppression of the HH+ peak in the SHJ focussing sample, we extract the scattering length of both spin peaks and compare this to the scattering length extracted from Shubnikov-de Haas oscillations. The amplitude of focussing peaks decays exponentially as the focussing path length is increased:

\begin{equation} \label{eq:ScatteringLength}
    R_{\text{Focus}} \propto A~e^{-\pi d/2l}
\end{equation}

where $l$ is the small angle scattering length \cite{spector_ballistic_1990,heremans_observation_1992,rendell_transverse_2015} and $d$ is the focussing diameter. By fitting a double Gaussian to the spin peaks for each focussing diameter, the amplitude of the peaks can be found as a function of path length.

\begin{figure}
    \centering
    \includegraphics[width=0.45\textwidth]{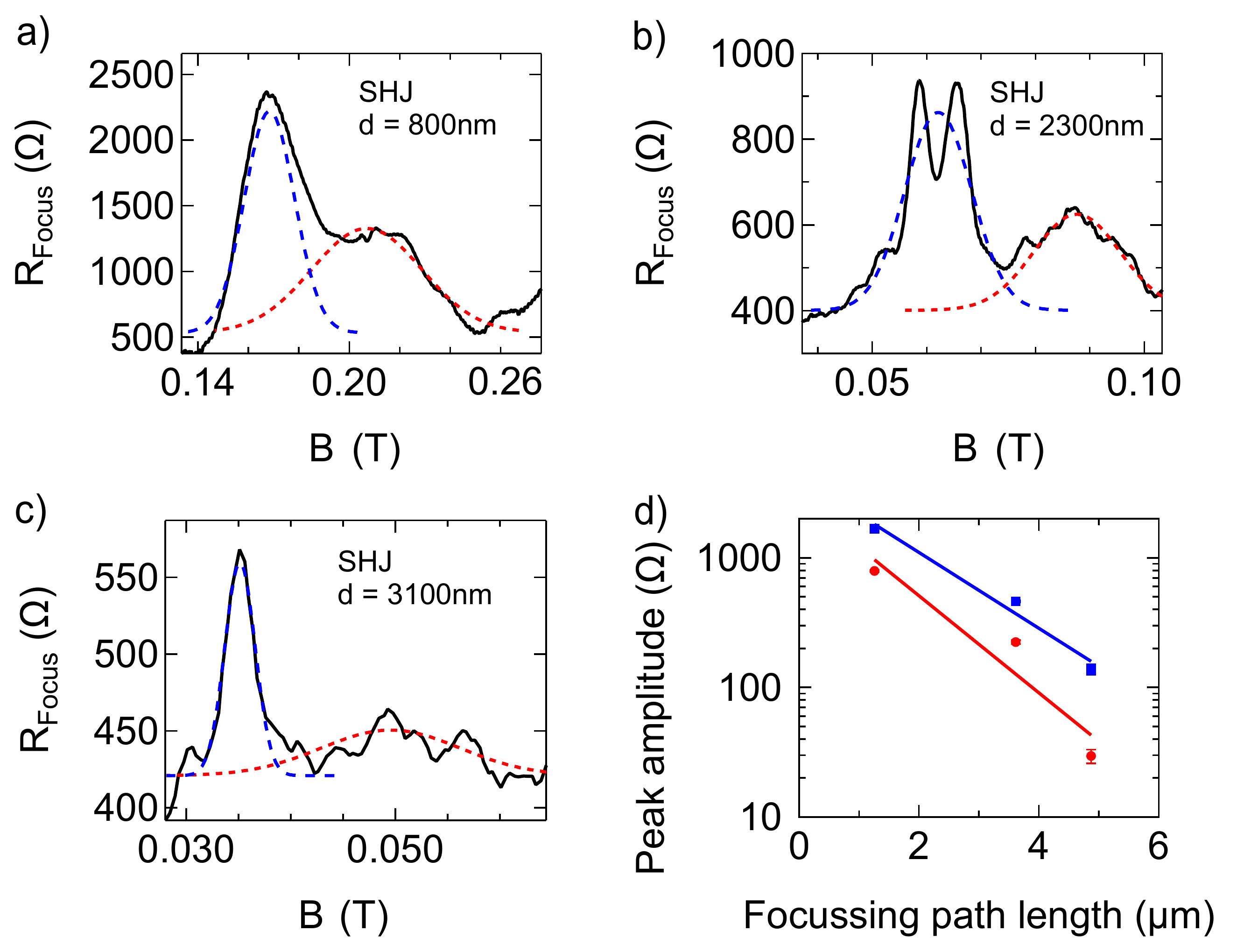}
    \caption{Focussing decay length for the SHJ sample. \textbf{a) - c)} Gaussian fits to the spin-split focussing peaks on the SHJ sample for all focussing diameters. \textbf{d)} fit to the amplitude of each spin peak with error bars showing the fit uncertainty. From this we can obtain the scattering length for each spin.}
    \label{fig:SHJFits}
\end{figure}

Fig \ref{fig:SHJFits} a) - c) shows the results of a double Gaussian fit to the SHJ focussing peaks for all focussing lengths. The amplitude of the spin split peaks as a function of focussing path length is plotted on semilog axes in Fig \ref{fig:SHJFits}d). A straight line fit to the data in Fig \ref{fig:SHJFits}d) allows the scattering length for each of the spin peaks to be found ($l_+$ and $l_-$). It is difficult to directly compare the scattering lengths found from focussing to a mean free path for two reasons. First, focussing measurements require the holes to travel through a narrow detector QPC, where even small scattering events can be sufficient to prevent a hole from reaching the detector. Second, the value of the scattering length found from focussing is sensitive to the choice of background resistance used for the peak fitting. However, the ratio of the scattering lengths is  independent of the background resistance. From the linear fits in Fig \ref{fig:SHJFits} d), Eq\ref{eq:ScatteringLength} can be used to find the scattering length for each spin state ($l_+$ and $l_-$). From this we find that $l_+/l_-$ = 0.77 $\pm$ 0.01.

Finally, the ratio of the scattering lengths is compared to a ratio found from Shubnikov-de Haas oscillations. The scattering length is given by $l \propto v_F\tau$, where $\tau$ is the scattering time. Assuming that $\tau$ is constant for both spin states, the ratio of $v_{\text{F}}$ (i.e. $v_+/v_-$) can be used to predict the ratio of $l_+/l_-$. The ratio of $v_+/v_-$ can be found from the ratio of $n_+/n_-$ since

\begin{equation*}
\frac{v_+}{v_-} = \frac{n_+}{m_+}\frac{m_-}{n_+} = \frac{k_+ m_-}{k_- m_+}
\end{equation*}

Assuming the subbands are approximately parabolic this expression can be simplified to 

\begin{equation*}
    \frac{l_+}{l_-} = \frac{v_+}{v_-} = \sqrt{\frac{n_-}{n_+}}
\end{equation*}

The values of $n_+$ and $n_-$ were found from measurements of the frequency of Shubnikov-de Haas oscillations on the same sample ($f = nh/e$). From this $n_+ = 1.21\times10^{11} \text{cm}^{-1}$ and $n_- = 0.68\times10^{11} \text{cm}^{-1}$ which gives a predicted ratio of $l_+/l_- = 0.75$, almost identical to the measured value of $l_+/l_- = 0.76$.

The good agreement between the predicted and measured ratios of $l$ indicates that the increased scattering of one spin state is the likely cause of the difference in peak amplitude observed in the SHJ focussing sample.

\section{\label{sec:Conclusions}Conclusions}

In this work we have investigated magnetic focussing with a large change in the magnitude of the Rashba spin-orbit interaction. We observed an attenuation of one of the spin peaks that is typically associated with a change in spin polarisation. Here we have instead shown that the difference in peak amplitude is consistent with a difference in effective mass and hence scattering rate between the spin chiralities. This result indicates that care must be taken when associating the amplitude of hole focussing peaks to spin polarisation in hole systems with $k^3$ spin-orbit interaction.

\section{Acknowledgements}

The authors would like to thank S. Bladwell, O.P. Sushkov, E.Y. Sherman and U. Z{\"u}licke for many valuable discussions. Devices were fabricated at the UNSW node of the Australian National Fabrication Facility (ANFF). This research was funded by the Australian Government through the Australian Research Council Discovery Project Scheme; Australian Research Council Centre of Excellence FLEET (project number CE170100039); and by the the UK Engineering and Physical Sciences Research Council (Grant No. EP/R029075/1).

\bibliography{ScatteringFocussingPaper}
\end{document}